\documentclass{article}
\usepackage{epsfig}

\def\Journal#1#2#3#4{{#1} {\bf #2}, #3 (#4)}

\def\NPA{{\em Nucl. Phys.} A}
\def\NPB{{\em Nucl. Phys.} B}

\def\PLB{{\em Phys. Lett.} B}

\def\PREV{\em Phys. Rev.}

\def\PRC{{\em Phys. Rev.} C}

\begin{document}

\def\DD{{\Delta \Delta}}
\def\DN{{\Delta N}}

\title{
Study of $\DD$ excitations in the reaction $n + p \rightarrow d + \pi\pi$
}

\author{C. A. Mosbacher and F. Osterfeld$^\dagger$ \\[0.2cm]
Institut f\"ur Kernphysik, Forschungszentrum J\"ulich GmbH,\\ 
D--52425 J\"ulich, Germany}

\date{Baryons '98, Bonn, Sept. 22--26, 1998}

\maketitle

\begin{abstract}
The deuteron spectrum in $n+p \rightarrow d+ \pi\pi$
at $k_n = 1.88$ GeV/c and $\theta_d = 0^o$
is explained by considering a $\Delta \Delta$ excitation as the reaction  
mechanism for the $2\pi$ production. We study the influence of intermediate $\DD$ 
and $\DN$ interactions on the spectrum. The angular distribution of the pions
is predicted.
\end{abstract}

\section{Introduction}

\begin{figure}[!h]
  \begin{center}
    \epsfig{file=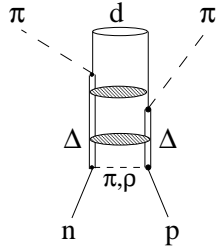, height=3.4cm}
    \caption{$\DD$ excitation mechanism in the reaction $n+p \to d+ \pi\pi$.
       The shaded areas symbolize the intermediate $\DD$ and $\DN$ interaction,
       respectively.}
    \label{fig1}
  \end{center}
\end{figure}

Experimental measurements of the reaction $n+p \to d+ \pi\pi$
at neutron momenta of 1.88 GeV/c \cite{plouin78}
show a typical cross section structure known as the ABC--effect \cite{abashian63}.
The main features of this effect can be explained by assuming 
a double $\Delta(1232)$ excitation to be the dominant 
reaction mechanism for the $2\pi$ production \cite{risser73,barnir75}. 
Therefore the $n+p \to d+ \pi\pi$ reaction allows us to study 
the influence of direct $\DD$ and $\DN$ interaction potentials.
The Feynman diagram corresponding to the $\DD$ mechanism is shown in fig.~\ref{fig1}.

\section{Theoretical Framework}

In order to incorporate the interaction dynamics of the 
$\DD$ system in a proper and efficient way, we apply a coupled channel 
approach \cite{mosbacher97} deduced from $\Delta$--hole models.
The matrix element 
is calculated using the so--called source function formalism \cite{udagawa94}.
We set up a system of coupled integro--differential equations for the correlated
$\DD$ wavefunction, which can be solved in configuration space
with the Lanczos method \cite{udagawa94}.
The $\Delta$ resonance is treated thereby as a quasi--particle 
with a given mass and an intrinsic, energy--dependent width. 
We construct the interaction potentials $V_{\DD}$ and $V_{\DN}$ 
within a meson exchange model \cite{machleidt87}.
The exchanged mesons taken into account are the pion ($\pi$), the rho ($\rho$), 
the omega ($\omega$), and the sigma ($\sigma$). All the parameters of the 
potentials, such as the coupling constants, cutoffs and meson masses,
are determined from other reaction studies, like single pion production
and charge exchange reactions on the deuteron \cite{mosbacher97}.

\section{Results and Discussion}

\begin{figure}[!t]
  \begin{center}
    \epsfig{file=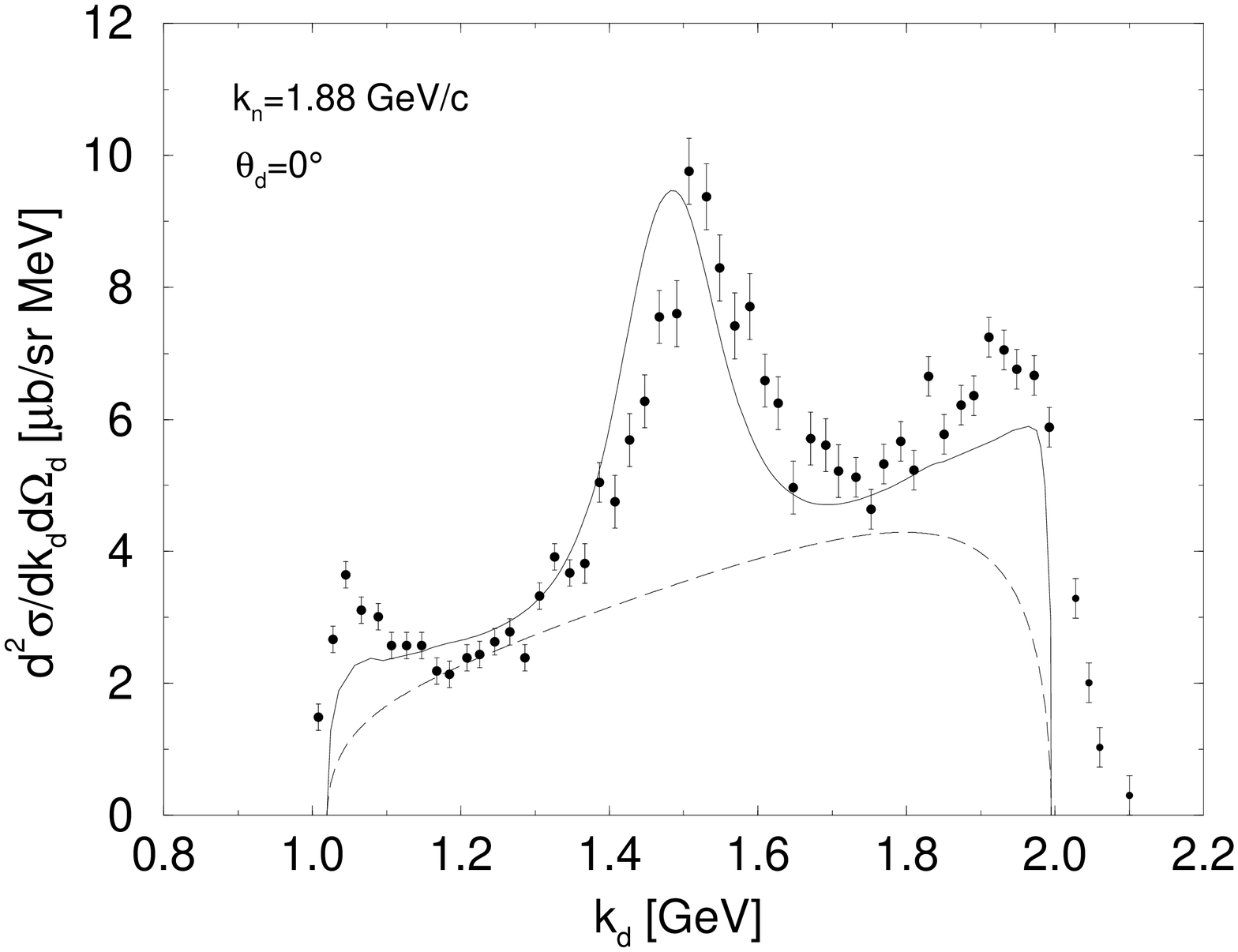, height=5.5cm}
    \caption{ Theoretical result for the $n+p \to d+ \pi \pi$ cross section 
      (full line) and experimental data \protect \cite{plouin78} 
      at $k_n =1.88$ GeV/c and $\theta_d = 0^o$.
      The dashed line shows the phase space.}
    \label{fig2}
    \vspace*{0.5cm}
    \epsfig{file=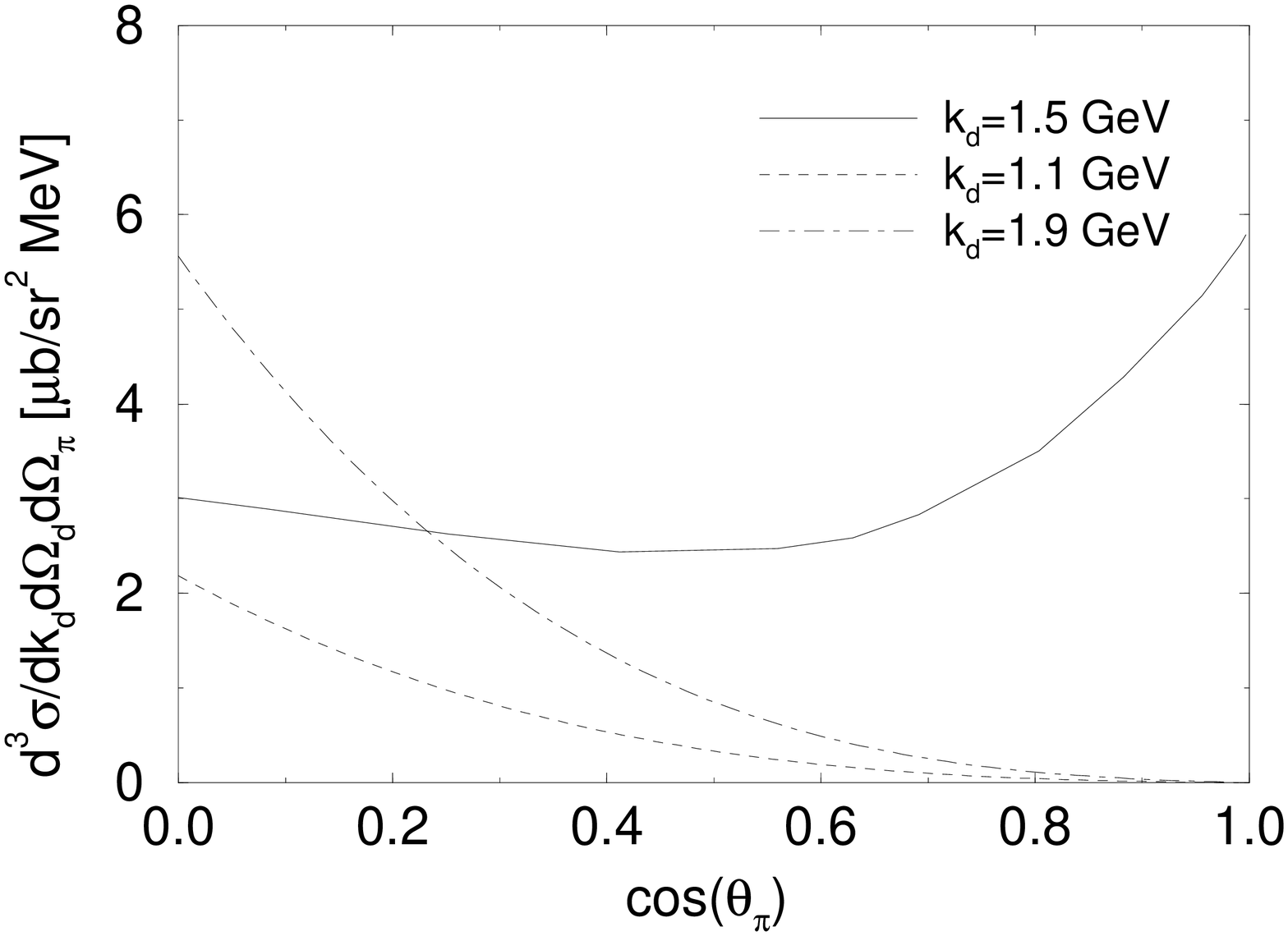, height=4.5cm}
    \caption{Angular distribution of the two pions in the deuteron rest frame,
      given as a function of $\cos (\theta_\pi) = \hat k \cdot \hat k_n$. }
    \label{fig3}
  \end{center}
\end{figure}

Fig.~\ref{fig2} shows the result of our full model calculation compared to the 
experimental deuteron spectrum for $\theta_d = 0^o$.  
The overall agreement is very good. 
The cross section exhibits characteristic 
enhancements at missing masses close to $2m_\pi$ (corresponding 
to maximal and minimal deuteron recoil momentum in the laboratory frame)
and at the largest missing masses possible (the central peak).
This structure can be easily
understood as follows \cite{risser73}. Let us define 
\begin{equation}
  K= k_1 + k_2 \, , \qquad 
  k= k_1 - k_2 \, ,
\end{equation}
where $k_1, k_2$ denote the four--momenta of the two 
pions. The $\DD$ mechanism is most efficient if both $\Delta$'s are 
on--mass shell, which requires 
\begin{equation}
  s_{\Delta_1} = \frac{1}{4} \: (k_d + K + k)^2 = 
  \frac{1}{4} \: (k_d + K - k)^2 = s_{\Delta_2} \, .
\end{equation}
This condition is equivalent to $k_d \cdot k = 0$ and 
can be satisfied in two ways:
\begin{enumerate}
\item $k = 0$ and hence $\sqrt{K^2} = M_{\pi\pi} = 2m_\pi$, or 
\item $\vec k_d = 0$ in the rest frame of the two pions.
\end{enumerate}
In the latter case, the $2\pi$ CMS is identical to the rest frame of 
the deuteron and hence the overall CMS. The two pions 
pick up the whole kinetic energy which means        
$M_{\pi\pi} =$ max. This leads to the high mass enhancement 
which corresponds to the anti--parallel decay of the two $\Delta$'s, 
while the low mass enhancement ($k = 0$) corresponds to the parallel decay.

These two different kinematical situations are also manifest 
in the angular distributions shown in fig.~\ref{fig3}. For $k_d = 1.5$ GeV/c (in the lab frame), 
the pions are emitted back to back and any angle between $\vec k$ and the beam 
axis is kinematically possible. The shape of the cross section reflects the spin--structure 
of the $\Delta \Delta$ excitation \cite{mosbacher97}. 
For $k_d =$ 1.1 and 1.9 GeV/c, the pions are emitted nearly parallel and have 
identical energies, hence $\vec k$ is dominantly perpendicular to the beam axis. 

\begin{figure}[!t]
  \begin{center}
    \leavevmode 
    \epsfig{file=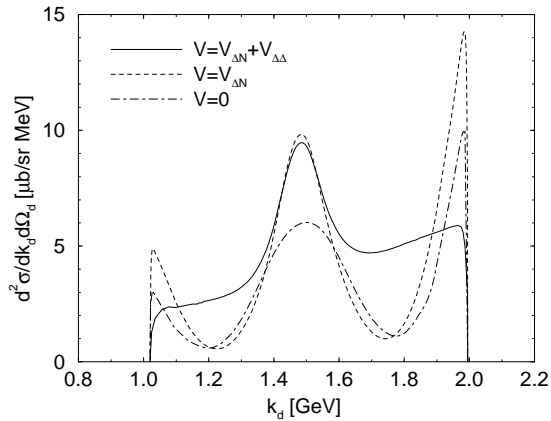, height=5.5cm}
    \caption{Influence of the $\Delta \Delta$ and $\Delta N$ interaction potentials.}
    \label{fig4}
  \end{center}
\end{figure}

Fig.~\ref{fig4} demonstrates the influence of the $\DN$ and $\DD$ interaction potentials
in our model calculation.
The result for $V=0$ (dashed--dotted line) clearly shows the three peak 
structure as discussed above. 
The potential $V_\DN$ is attractive and thus results in even more pronounced 
peaks (dashed line) since it simply lowers the overall excitation energy.
Inclusion of $V_\DD$, however, leads to a 
redistribution of strength from the peaks towards the kinematical 
less favored situations between (solid line). This is the case because the optimal
configuration with both $\Delta$'s on mass-shell can now also be reached 
if the initial momentum distribution was asymmetric.

To summarize we have shown that
1.\ the $\Delta \Delta$ excitation is the 
dominant reaction mechanism in the $np \to d + \pi\pi$ two pion 
production at $k_n = 1.88$ GeV/c, and  
2.\ the intermediate $\DD$ and $\DN$ interactions play an 
important role in this reaction which therefore may serve as a tool for
closer examination of the interaction potentials.    

This work was supported in part by the Studienstiftung des deutschen Vol\-kes.
We are very grateful to C. Hanhart and J. Speth for many helpful discussions.


\end{document}